# Bridging the gap: Generating a design space model of Socially Assistive Robots (SARs) for Older Adults using Participatory Design (PD)

Adi Bulgaro* (ORCID 0000-0001-6188-4389) Ela Liberman-Pincu (ORCID 0000-0002-9753-7714), Tal Oron-Gilad (ORCID 0000-0002-9523-0161)

Ben-Gurion University of the Negev, Industrial Engineering and Management, Be'er-Sheva, Israel.

*Corresponding author: adibul@post.bgu.ac.il

Contributing authors: orontal@bgu.ac.il, elapin@post.bgu.ac.il

**Abstract**

Participatory Design (PD) methods are effective in understanding older adults' perspectives, concerns, and wishes and generating ideas for new intelligent aids. The aim of our study was first to map perceptions and explore the needs of older users from socially assistive robots (SARs) and then, to integrate new tools and experiences for end users to express their needs as part of a PD process. The outcome of the process is a design space model of functional, behavioral, and visual relationships and elicited emotions. This process enables to explore, map, and understand the needs of older users before and after experiencing with SARs, and to learn how they impact robotic designs, behavior, and functionality. First, by interviewing older adults, caregivers, and relatives we learned older adults' daily routines, habits, and wishes. Then, we reconstructed those needs into design requirements and further detailed them with older adults using focus groups. Based on the functional, behavioral, and visual design factors that emerged from this phase, we built experimental human-robot tasks, on a commercially available robot, to examine feasibility and acceptance of the technology in one-on-one interactions. Participants' responses throughout the study led to the creation of the design space model mapping relationships and elicited emotions.

**Statement and declarations.** Students were supported by scholarships, as noted in the acknowledgments.

**Acknowledgments.** This work was partially supported by The Israeli Ministry of Innovation Science and Technology (MOST; grant 3-15625), by the Ben-Gurion University of the Negev through the Helmsley Charitable Trust, the Agricultural, Biological, and Cognitive Robotics Initiative, and by the George Shrut Chair in Human performance management.

# 1 Introduction

Robots for elder care is an evolving field gaining momentum [1]. The hope is that future robots will complement and support the caregiver workforce as the increase in the older adult population will be accompanied by requirements for more elder care [2]. Already in 2008, nearly half of the older adult population relied on outside help. In the upcoming years, the demand for elder care may reach levels at which traditional care will not be able to cope with [3]. Solutions are necessary to overcome the anticipated shortage of caregivers and allow older adults age in place [4].

Assistive robots can support older adults in daily activities [5]. However, current adoption is limited, and many robotic systems are not fit for eldercare. We already know that acceptance of robots depends highly on the quality of interaction [6] affecting users' perception and experience [7]. Further, mass customization (a process in which the user actively modifies aspects of a product by selecting predefined modules) can positively affect users' acceptance of a SAR [8]. In an older study, Broadbent et al. [9] looked at older adults' responses to healthcare robots and found that it is possible to increase their acceptance of by correctly evaluating their needs. Cooper et al. [4] agree that new assistive technologies are required to meet older adults' needs, including finding solutions to support and expand independent living at home. At the same time, they argue that the existing deployment of assistive robots, is still rare. In Olatunji et al. [10], the authors point out particular requirements for future mobile and manipulator robots to be used in older adults' homes (see also [11]).Wilkinson and De Angelli [12] emphasis the importance of including older adults in the design process of systems for older users. They warn of the 'Cycle of design oversight' where designers fail to recognize older people's needs and fail to engage with potential users. Via their design use cases for an intelligent mobility aid and a wheelchair, they showcase how adopting a variety of participatory design (PD) techniques is beneficial to the design of a new product noting that older adult users were quick to suggest ideas for new product developments while focusing on functionality, behaviors, empowerment, and aesthetics.

PD is a design approach characterized by user involvement and is part of adopting a use-(or user)-centered design approach to contribute to setting requirements and improving the integration of robotic systems for older users [13]. PD techniques are useful to extract user needs, identify the robots' essential characteristics and set meaningful use cases [14]. Yet only few studies on SARs for older users utilized PD methods. Rogers et al. [15] argue that applying different methods of PD with diversion of the participating older adults is effective in understanding the elderly's perspectives, concerns, and wishes. Yasuoka and Kamihira [16] describe an extended PD process model with an emphasis on collaborations; Professional: collaboration of designers and developers, Creative: including multi-stakeholders from different backgrounds; User research involving users and designers and learning by doing; observations of real interactions in realistic settings). Hence the challenge of developing SARs for older adults begins with hard rigorous work of how to extract older adults' insights and desires, include and involve other stakeholders in eldercare (e.g., family caregivers or professional caregivers) and create experiences. Indeed, developers of healthcare robots tend to use a bottom-up approach in their design process, where they first start by defining features of the interface and then map them for a possible implementation [17], [18]. Yet, often when considering the visual and interaction design of the robot, it is perceived as something that can be adapted later to the personality of the designated users, their needs, and personal taste [19], since the older population is diverse, and their needs change dramatically with age. Further, designers should consider common needs of older adults together with requirements of stakeholders [20]. Recently, Bradwell et al. [21] provided design recommendations for socially assistive robots for health and social care based on a group activity with 232 stakeholders, mainly healthcare professionals but also healthcare students and other professionals. Utilizing eight lunches and exhibitions of robots followed by round table discussions, commercially available SARs were available for participants to approach, engage and discuss. Their results revealed that key stakeholders were open towards using SARs. Furthermore, they identified potential uses, indicating the possibilities they found for SARs. At the same time, they proposed design improvements to ensure usefulness. Specifically, they noted the need for improved mobility for uneven floors, improved voice recognition, better ease of use, autonomous charging, soft and friendly aesthetics, non-robotic design to improve friendliness and androgynous appearance. The authors claim that a physical demonstration of various types of SARs is the strength of their study, allowing more comprehensive attitude formation from participants than focusing on one type of SAR. Examining multiple SARs in a focus group [21] is an addition beyond the studies that have been done so far, but there is still another dimension that is missing. Older adult participants were not actively participating in the design but rather grading existing designs. Hence, participants could not fully express their feelings beyond the discussion about the existing designs. Another step is still missing. Notably, studies involving older adults often do not provide means to express their thoughts and preferences in their own language and priorities [22].

The aim of the current study was first to map perceptions and explore the needs of older users from socially assistive robots (SARs) and then, to integrate new tools and experiences for end users to express their needs as part of a PD process towards creating a design space model of functional, behavioral, and visual relationships and elicited emotions. To do so, we followed Rogers et al. [5] three key phases for participatory design (PD) in human-robot interactions (HRI) and conducted **interviews** with older adults, caregivers, and relatives, then **focus groups** with older adults, and finally **one-on-one HRI experiences** with a commercially available temi robot. First, by interviewing older adults, caregivers, and relatives we learned older adults' daily routines, habits, and wishes. Then, we reconstructed those needs into design requirements and further detailed them with older adults in the focus groups. Based on the functional, behavioral, and visual design factors that emerged, we built experimental human-robot tasks, on the temi robot, to examine the feasibility and acceptance of the technology and the consequent elicited emotions. Throughout the various stages of the work, we formed the design space model to enable better understanding of the meaning of the socially assistive robot in the eyes of older adults. We characterized the building blocks of the design space model through understanding and mapping older adults' needs, perceptions and preferences regarding SARs' functionality, behavior, and appearance.

## 2 Methodology
## 2.1 Overview

The work was conducted as a progressing process. From each step we designed the following one according to the results obtained (see Figure 1). The process started from mapping of needs of the older adults throughout the day via interviews. Following, we continued with focus groups that shed light on functional, behavioral, and visual appearance aspects of SARs for older adults. Using the outcomes, we developed an experimental setting with the temi robot addressing functional, behavioral, and visual design aspects and conducted One-on-one experiences (see Table 1).

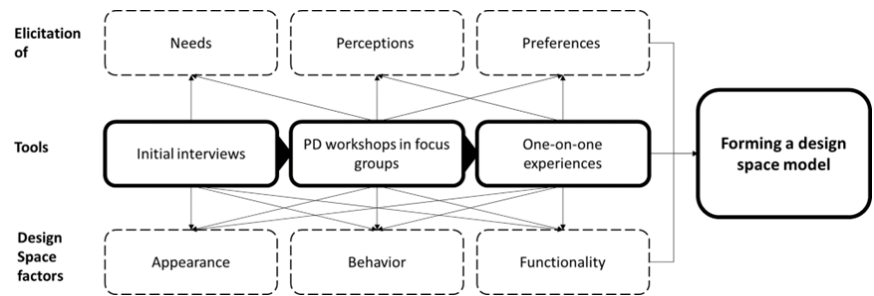

**Figure 1** Research process

**Table 1** Components of the experimental tools

|  | **Initial interviews** | **PD workshops in focus groups** | **One-on-one experiences** |
|---|---|---|---|
| **Robotic context** | Theoretical, no specific robot, focus on daily activities and routines of the older adults | Existing robots ("off the shelf") in three contexts medical, household, and social Then future design of a non-existing robot (wishes) | The robot as a personal assistant in the home of the older adult user. |
| **Size of population** | 24 one-on-one interviews with older adults (14), relatives (5), and caregivers (5) | Two groups of five older adults living alone | 31 older adults living in an assisted living facility "Palace Lehavim" |
| **Focus** | Older adults' needs, daily activities | Perception of robots in context (via existing robots) and in general in the future (wishes and expectations) | Real interaction. Gaining experience with the robot's functionality, behavior and look. |
| **Tools** | None | Printed cards | Robot temi |
| **Functional** | Assessing the functional needs of older adults throughout the day | Discussing use contexts in which a robot can help | Evaluating navigation and cognitive tasks |
| **Behavioral** | Understanding older adults' daily routines and wishes | Discussing future human-robot relationships | Assign humor to the robot to establish a friendly relationship with the user |
| **Design** | Exploring existing designs that participants are familiar with | Participatory design process- Each group designed one "ideal" robot | Allow design adjustment to an existing robot |

## 2.2 The robot

The robot we used in the practical experiment is temi (www.robotemi.com). For the various tasks and their examination in each of the factors, we made adaptations to the capabilities available at temi and developed customized applications. temi's development environment is in "Android Studio" in Java. It has a control center called "temi center" where remote operations can be performed and other features of the robot that can be done by using that center. In addition, we used temi's video calls as part of the robot's app that can be downloaded from the app store.

# 3 Initial interviews

The interviews were aimed at understanding the desired relationship/s between the older users and the robot, what interaction the user aspires to reach when using a robot and how the other stakeholders foresee this interaction. See also Table 1.

## 3.1 Method for interviews
### 3.1.1. Participants

Interviewees were recruited independently by contacting older adults with the necessary profile – living alone and over the age of 70, by telephone. The recruitment was made through our relatives and acquaintances who connected us to potential interviewees. Also, we recruited interviewees from assisted-living homes and directly contacted neighbors living nearby. There were no connections between and among the participants of the three groups, but it is possible that in the group of the older adults some were acquainted due to the proximity of living or hanging in common places.

*Older adults.* Fourteen interviewees (four males and ten females) from the city of Beer-Sheva in the south region of Israel, aged 71-85 participated. All interviewees were older adults who live alone. Three older adults live in an assisted living facility, and eleven in their own homes.

*Relatives.* The relative interviewees included five people (three men and two women) who have older parents that live alone, aged 48-57, also from Beer-Sheva. These interviewees were not related to the participants from the older adult group.

*Professional Caregivers.* The caregivers group included five professional caregivers (four women and one man) aged 38-50, three who came from foreign countries and two who live and work in a daycare facility for older adults in Beer-Sheva.

### 3.1.2. The interview processes

The interviews were conducted in the spring of 2021, in a comfortable environment for the interviewees - at home or in the assisted living home. We built an interview guide that contained the same question base for all interviews. We conducted the interviews by going through the questions from the interview guide. Whereas, at the beginning of every interview, each participant was presented with the subject of the research. The older adult participants were asked to specify their daily routine, and the interviewer asked questions that helped focus the interview on the necessary information. Following these questions, the older adult participants were asked about their difficulties and challenges. The closing questions of the interview combined questions about the possibility of integrating robots into older adult participant's life. The differences between the interviews of the older adults and the other two groups were reflected in questions aimed at understanding the difficulty and challenges they see for older adult and what they feel could have helped the older adults they know. An interview lasted between 40 minutes to an hour.

### 3.1.3. Thematic analysis of the interviews

The data derived from the interviews were analyzed using thematic analysis [23]. Thematic analysis analyzes qualitative data by which ideas, common topics or patterns can be extracted from the text. According to [24], thematic analysis is a method rather than a methodology and therefore it is very flexible. We used the necessary phases from the process of thematic analysis. First, we re-read the interviews to get to know them on a deeper level. Then we coded the interviews by finding repeating sentences or meanings and highlighting them with colors to create a legend of relevant titles. Following, we were able to identify the main themes and the subjects in the text that were related to each of them. Lastly, we used word clouds to detect the most frequent words [25].

## 3.2 Interview Outcomes and Results
### 3.2.1. Differences between stakeholder groups

The interviews raised insights into the perceived differences in needs among the interviewed groups. From the <u>*caregivers'*</u> perspective, older adults need someone who can listen to them, some daily activity that will keep them busy, cognitive support, medical support, and close supervision. Regarding the <u>*relatives*</u>, they presented requirements for *medical help*, such as supervision and monitoring the older adult. Most of the relatives who participated presented an opinion about *reducing loneliness*, they implied that their older family member is alone during the day for long periods of time. They were concerned about the *medical help* and *physical help*, i.e., whether their parent was functioning well. In the same issue, they were concerned about scenarios where their parent is falling at home, and if they're not taking their medications, etc. Their requirements, to help with those issues, included first aid in case of need, receiving vital measures from their parent, entertaining them during the day, and so on. As far as the <u>*older adults*</u>, most of them presented a need for *help with household chores and tasks*; cleaning and cooking and carrying groceries from the store. Also, *physical support*, such as supporting in arising when sitting or lying on the bed and assisting in the shower. In addition, they wanted *reminders* for tasks, events, or objects they forgot where they put, as well as social help. Hence, there are many requirement differences among the groups.

Dividing the design requirements into physical and social requirements, we can delve deeper into this subdivision to understand how many participants presented each type of need (see Table 2 ). Physical needs were more apparent in the interviews with older adult participants, and very few of them expressed social needs. On the other hand, most caregivers presented both physical and social needs of older adults from their point of view.

**Table 2**  The division into stakeholder groups by needs and requirements.

|  | Needs | Older adults (n=14) | Relatives (n=5) | Caregivers (n=5) |
|---|---|---|---|---|
| **Physical needs** | Physical support | 10 | 3 | 3 |
|  | Help in daily chores | 10 | 2 | 0 |
|  | Reminders | 5 | 1 | 1 |
|  | Medical help | 7 | 4 | 4 |
| **Social needs** | Reducing loneliness | 4 | 1 | 3 |
|  | Communication | 4 | 2 | 4 |

### 3.2.2. Gender differences

We do not have a large sample of interviewees, but it was noted that the words that were repeated most in the interviews with the older adults reflected physical and social needs (as noted in 3.2.1). The words that came up most were interaction – talk - conversation (16 times), help (10 times), love (9 times). The word clouds portray some differences between men and women in needs and perspectives (see Figure 2.a; Figure 2.b). For example, when interviewing women, the words that came up most frequently called for warmth and love (9 times), and the most prominent words were related to family (sister, family, kids, 14 times), talk (8 times). In contrast, when interviewing men, the words that came up most frequently called for action and therefore the most highlighted words were outside-go-active-play (10 times).

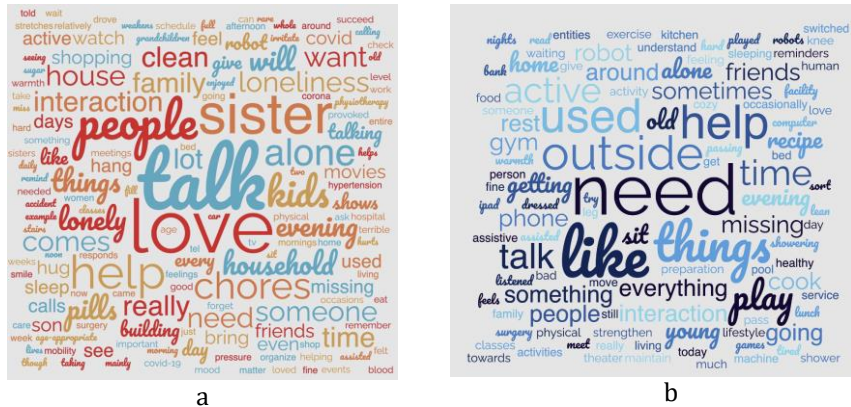
**Figure 2** Clouds of words: (a) Women's needs, (b) Men's needs.

### 3.2.3. Summary of interviews results

Using the outcomes of the thematic analysis we divided the interview outcomes into themes. Two main themes emerged. The first, relates to the daily needs of older adults who live alone and is comprised of 6 sub-themes: (1) Needs, (2) Unresolved needs, (3) Solutions the older adults found, (4) Needs from the past, (5) Needs following medical treatment, (6) Lack of the need to get help. The second relates to older adults' attitude towards technology and is comprised of five sub-themes: (1) Fears of robots, (2) Fears the robot should moderate, (3) Criticism of using technology, (4) Difficulties past, (5) Finding solutions for others or the future. The detailed outcomes from the thematic analysis are summarized in Appendix A (Table 10, for daily needs, and Table 11 for attitude towards technology).

## 4 Participatory Design workshop in focus groups

The PD workshop focus group's purpose was to gain further understanding of the needs of older adults by detailed examination of existing SARs and utilize our PD tool to create new concepts for robots.

### 4.1 Method for PD workshop focus groups
#### 4.1.1. Overview

The PD focus groups the group discussion began with needs and wishes for a context of use (medical, household, and social), then looking at existing "off the shelf" robots for each aspect (functional, behavioral, and visual design). Following participants were asked to design their own robot. This "hands-on" experiences utilized the toolkit [11]. The toolkit consists of physical prints of parts of robots built from a market survey (e.g., various body shapes, screens, wheels in varying color schemes) see [8], [26]. As their final outcome of the focus group, in each focus group, the participants created their ideal robot using a pre-prepared toolkit of robot parts (see Figure 3). Insights from the first focus group were implemented to improve the procedure of the second focus group. Specifically, we added more guided questions following the design of the robot with the Toolkit, to better understand the outcomes of the design and allow the participants to explain their choices.

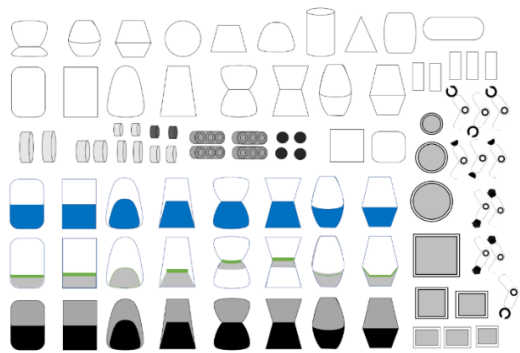
**Figure 3** Two-dimensional robotic parts in the toolkit.

### 4.1.2. Participants

Ten participants, living alone aged 75-80, participated in the first and second focus groups (3 women and 2 men in each). Both focus groups included a sample of participants who live alone in the southern region of Israel. The participants recruited independently by contacting older adults with the necessary profile by telephone. The recruitment was made through our relatives and acquaintances who connected us to potential interviewees, therefore they represent a convenience sample.

### 4.1.3. Process and tools

All five participants sat together with the experimenter and had a discussion with the help of the preprepared interview guide. A PD focus group meeting lasted 80 minutes.

In the beginning, participants were presented with the purpose of the discussion. The experimenter used an interview guide that included partition to three contexts of use that arose significantly from the interviews: household, healthcare, and social robots (see Table2 ); the discussion was divided into these three contexts (see Table 3). For each context, the participants were asked to describe the features that a robot in the field needs to convey. They were presented with a list of features (responsible, reliable, likeable, friendly, other etc.) and needed to decide what features best suit a robot in the relevant context of use. Then the participants described suitable visual and physical features. Following the discussion, they were asked to find a robot, from a robot repository, that best represents the features brought up (see Figure 4; Figure 5; Figure 6). The robot repository included 4-5 photos of robots that exist in the market today and can assist in the specific activity for every context. This process was repeated for the three contexts of use. In the last stage of the focus group, participants were asked to create their own robot design using a toolkit we made for them. The Toolkit included physical prints of parts of robots built from a market survey conducted prior to this study, which examined the visual characteristics of existing robots, and thus the components were built in favor of the focus group [8], [26]. The parts represented the preferred shapes, screens, wheels, and colors that matched the results from the survey (see Figure 3).

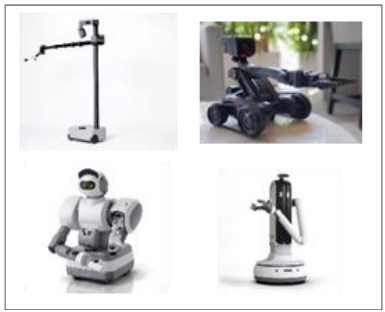

**Figure 5** Household robots: Stretch RE1, Nobot AI, Aeolus Roas, Bot handy.

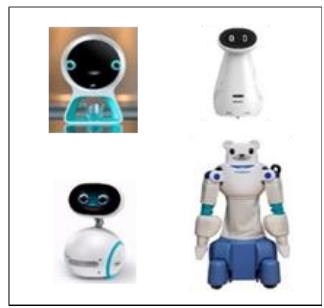

**Figure 6** Healthcare robots: Pillo, Bot care, Zenbo, Robear.

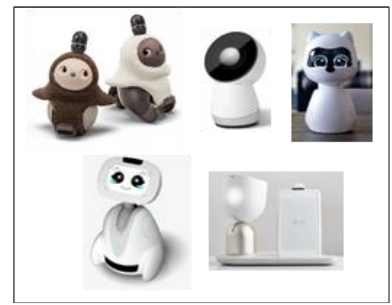

**Figure 4** Social robots: LOVOT, Jibo, Kiki, buddy, ElliQ.

**Table 3** Flow of a focus group

| Section 1: Help with household chores | |
|---|---|
| **Subjects from the process** | **Participants response** |
| A. **Activities that convey the characteristics** | The robot needs to contact us by first name and make a thorough job with responsibility. The robot should speak pleasantly, should update about what it is going to do next. |
| B. **The connection with the user** | We need to be able to talk with the robot and that it will respond accordantly, a good communication is important, also an option of dialogue is good. |
| C. **Location at the house** | The robot needs to stand somewhere in the space. It should not run after us; it should stay in one place. |
| **Section 2: Medical help** | |
| **Subjects from the process** | **Participants response** |
| A. **Activities that convey the characteristics** | The robot should ring an ambulance, have first aid, a defibrillator and get out of the house to get help. |
| B. **The connection with the user** | The robot should communicate and ask how you feel, pay attention to the vitals, and communicate with the necessary people. |
| C. **Location at the house** | The robot should be mobile, walk all day and follow us. It should turn around the house to check on us. |
| **Section 3: Social help** | |
| **Subjects from the process** | **Participants response** |
| A. **Activities that convey the characteristics** | The robot should read me the paper, have knowledge on wide information. I want to be able to discuss with it on anything. It should be able to play music, turn the TV and call my family with video. |
| B. **The connection with the user** | The robot must communicate, create a dialog |
| C. **Location at the house** | The robot should stay in one place most of the time and from time to time to check on me. |

## 4.2  PD Focus Group Outcomes and Results
### 4.2.1. Selected robots

At the end of each section in the focus group, the group was asked to select one commercial robot suitable, in their opinion, for the context of use and the features that were raised in the group discussion (see Table 4). Samsung's "Bot Handy" robot was chosen to help with home chores in both focus groups for household chores. According to the participants, the robot was chosen because its base looks stable and the ratio of body to base looks good. They thought it could help well at home because of its hand that allows it to perform daily tasks. It didn't seem too big to them so it wouldn't take too much space but also not too small so it can help in a good way. They also mentioned that this robot is close to what they imagined.

In the context of medical help, there was unanimity between the two groups selecting the robot. In the first group, the chosen robot was the "Robear" which is an experimental medical robot. This robot is designed in the shape of a bear and was developed medically to help older adults in homes. It's designed to lift people, carry them out of bed, and help in standing up. The robot was chosen because it appeared as a strong and stable robot that could assist if necessary. It also seemed authoritative to them because of its size and one that could perform various tasks. On the other hand, the second focus group chose "Zenbo" as their medical robot. This robot has basic medical monitoring functions such as measuring physiological. When asked to explain why they chose this robot, they explained they connected to it most and appreciated its movement ability, which they think is important when considering a robot made for medical purposes. There was still a reference to the "Robear" robot in the second focus group; the participants mentioned that they wouldn't pick this robot due to its oversized dimensions.

In the context of social support, both focus groups selected "buddy", a sociability robot that aims to win the hearts of its users. It is designed to present a wide range of emotions and convey it by interacting with the user. The participants chose this robot because it seemed cute, nice-looking, sociable, small, and didn't seem to take much space.

**Table 4** Selected robot by field for each focus group.

| Focus group \ Context of use | Household chores | Medical help | Social help |
|---|---|---|---|
| First focus group | 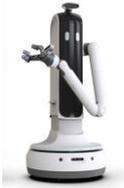 | 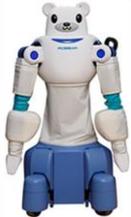 | 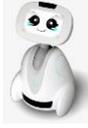 |
| Second focus group | | 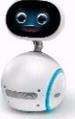 | |

### 4.2.2. Designing their "Ideal" robot - the first focus group

When asked to create their own robot design, participants decided to create a robot using a dark color scheme, in contrast to their declared preferred color, white. The results showed that there's a gap between what people think they need and what they want. Nevertheless, they did add eyes to the screen and created symmetry of two arms, so it did match the additional preferences that they presented in section 4.2.4 (see Figure 7).

### 4.2.3. Designing their "Ideal "robot - the second focus group

When the second group built their robot, they chose a white and blue color scheme as their declared preference of color in the discussions. They explained their choice by saying that blue indicates reliability. They noted that in their opinion, the robot should combine all three contexts of use but at the same time asked to make a simple robot without many functions that will make it difficult to manage. One participant noted that it may still be worth incorporating many functions, thus making the grandchildren come more often to help with the use of the robot. When considering the appearance and functionality, the choice of different hands was premeditated. They sought to design two different hands, one hand was the active one (i.e., the hand that will do the action itself) and the other hand, that is shorter will be used to assist and support the active hand (For example, the active hand will pick an object, and the supporting hand will support holding it while the robot moves). When looking at the legs, they were placed in a way that could show the mobility of the robot. Different eye colors were suggested emphasizing that the robot sees the world optimistically with "pink eyes" and performs actions with the other eye. The distance between the eyes indicates the ability to see in all directions (see Figure 8).

### 4.2.4. Comparisons between outcomes of the two focus groups

When comparing the outcomes of the two groups, it's possible to see differences but also some similarities. A noticeable difference is in the robot's colors, where the first group chose a dark color scheme, and the second group chose a white and blue combination. But there are still some identical components; Both robots have a distinct face and a buffer between the head and the body. When asked what the purpose of the buffer was, participants explained it was a bow tie to make the robot cuter. In addition, the two robots have two hands and a component (wheels or legs) that indicates mobility.

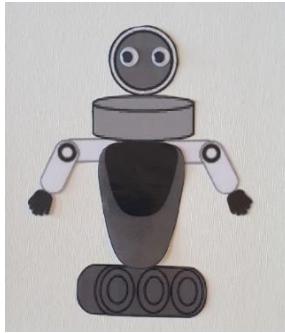

**Figure 7** The first focus group's "ideal" robot. Note the two arms and the wheels.

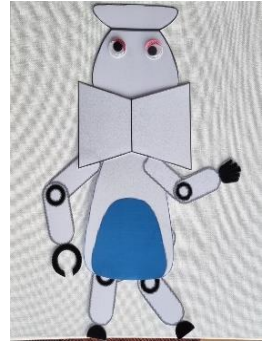

**Figure 8** The second focus group "ideal" robot. Note the left eye is pink and the right is "active". One arm has a gripper "active hand" and the other is smaller and used to support the active hand.

# 5 One-on-one HRI experiences

The purpose of this part was to examine the three aspects found in the previous sections: functionality, behavior and appearance. To do so, we developed an experimental setting addressing functional, behavioral and visual aspects using the temi robot. The experiment was conducted in the "Palace Lehavim" assisted living in a public room allocated specifically for the purpose of the experiments. The room was not in the private rooms of the residents but rather in the public environment that familiar to them. During the experiment, a single participant was in the room with the experiment instructors.

## 5.1 Experimental tasks for the functional domain

Using the existing interface of the temi robot and adaptations, we developed two tasks and included them in the interface, as described in Table 5.

**Table 5** The Functional domain tasks

| | | **Functional** |
|---|---|---|
| **Task 1: remote controlling** | Purpose | To examine whether older users will be able to use the robot to navigate in the house. With the help of remote control, the robot will be able to help identify hazards while maintaining the sense of independence of older adults. |
| | The task | To test the possibility of mobility in the house using a robot, we used the video calling option available in the robot. During the experiment, the robot's app (downloaded from the app store) was installed on a tablet. Participants directed temi to a picture located in a distant spot they could not see from their position. Using the arrows that appear on the screen during the video call, they could navigate the robot to the required position. |
| **Task 2: cognitive games** | Purpose | Using the cognitive game task, we wanted to test whether older users feel that a robot can help them maintain and improve their cognitive abilities, despite the technological limitations that older users usually feel. |
| | The task | To examine the possibility of helping older users in the cognitive aspect, we developed cognitive games. We installed these games on the robot and displayed them on the screen. The participants could choose to play one or several games within the allotted time frame for the task. |

### 5.1.1. Navigation and cognitive tasks

For examining the cognitive field, we developed and adjusted several applications that we presented on the robot's screen in a convenient way for users. In the navigation task, in which we wanted to examine the physical aspect, we used the navigation capabilities of the robot along with the video call option available in the robot. Using a tablet and temi's motion control operation and camera, the participants could lead temi, without moving from their chair to the designated place. We connected the users so that they could control the robot remotely and complete the navigation task. The robot was designed to be familiar with the experimental room, marking fixed places such as the starting point and the endpoint of the experiment (see Figure 9).

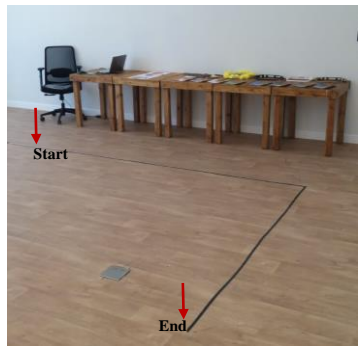

**Figure 9** The room used for the experimental sessions at the assisted living facility. The participant sat on the chair, the design options were laid on the tables, and the navigation path is marked by a black line on the floor.

## 5.2 The behavioral domain

We built structured conversations with the temi robot, as presented in Table 6. We gave the robot a nice and friendly personality by adding humor to the process to make the participant connect and feel comfortable with it.

**Table 6** The Behavioral domain explanation

| | | **Behavioral** |
|---|---|---|
| **Starting conversation** | The process | For the opening conversation, we created a sequence in the temi center to which we added three speech acts along with waiting time for the participant's answer. At first, as part of the first action of the sequence, temi greeted the participant and indicated that it was good to be here and asked the participant for his name. temi waited for an answer and then moved to the next action that included a sentence that indicated that it was pleasant to get to know the user and asked for his age. In the third speaking action temi added a joke and said that the participant did not look over the age of 25. |
| **Ending conversation** | The process | For the end conversation, we built a sequence to which we added two speech actions. In the first part, the robot noted that we had finished the experiment, praised the participant, and asked if he would like to meet again. After temi waited for the participant's answer he said he had a great time and greeted the participant goodbye. |
| **Purpose** | | The purpose of the conversations with the robot along with the addition of humor was to examine whether the participants enjoyed the interaction with the robot, whether adding humor perceived as positive and if it adds to the feeling of closeness to the robot, and to examine their actions during the conversations. |

### 5.2.1. Verbal connection with the robot

To make the conversations in the process adapted to the experiment, we used the sequence option located in temi Center. There were several conversations during the experiment. In the starting conversation, we wanted to make an acquaintance between the robot and the user, therefore the questions were general but with a joke to "break the ice". During the experiment, we added a few sentences that the robot said to create a flow in the experiment and let the robot show the actions to be performed. At the end of the experiment, we combined a concluding conversation in which the robot summarized the meeting and asked about additional meetings.

## 5.3 The visual design domain

Table 7 shows the process in the design domain. As part of the visual design factor, some of the participants had the option to customize and add visual elements to temi (see Figure 10).

Table 7 The Visual design domain explanation

| | | Visual Design |
|---|---|---|
| **Design phase** | The process | In the design process, we built 4-character options of heads of temi, with which we let the participants choose whether they wanted to create a child, dog (two colors), or flowery robot. |
| | Child head | Some of the participants who chose the child's head implied at the need for a friend, someone to talk to. For example, "What am I going to talk to a dog?". There were also those who designed based on personal taste and chose this head because of the color. |
| | Dog head | Some of the participants who chose the dog implied that they wanted something light. For example, "A dog is nice because it's like a pet without having to be cared for", while some participants were embarrassed by the choice, considered it as a game or something amusing and giggled while choosing that head. |
| | Flowery head | The participants who chose the floral head wanted to "get out of the box". There were those who chose it because it looked nice in their opinion, and also flower lovers chose it. The choice was made mainly not by context but by personal taste. |
| **Purpose** | | By using participatory design, we wanted to examine the relationships that are created between the participant and the robot. By choosing from three possible characters, we wanted to examine the impact of the design on the course of the experiment. |

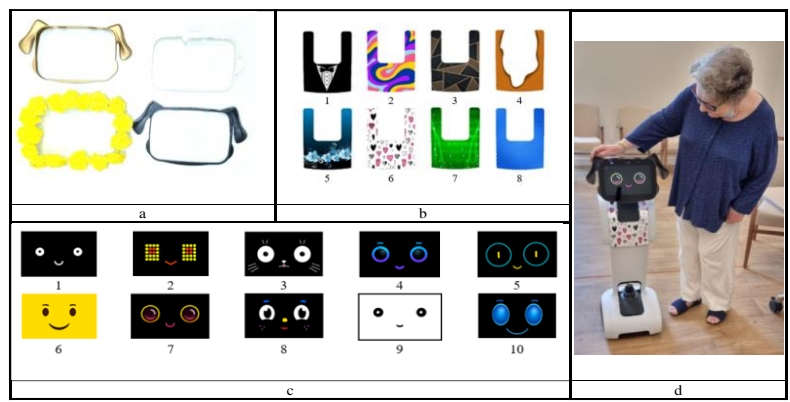

**Figure 10** a. External part of the screen; b. Vests for temi's chest; c. Faces for temi's screen as presented to the participants in the experiment; d. An example of participant with her choices of design.

## 5.4 Method for the experiment with the temi SAR
### 5.4.1. Overview

While interacting with a real robot, older adult participants performed tasks related to functional, behavioral, and visual design aspects. Using the existing interface of the temi robot with adaptations, we developed two functional tasks: navigating the robot indoors and using it for cognitive practice (see Table 5), built structured conversation capabilities (see Table 6), and designed a set of add-ons to affect temi's visual appearance (see Table 7). Our goal was to examine the participants' reaction and from that extract the elicited emotions of older adults. We designed an experiment to gain understanding of the functional, behavioral, and visual design preferences for SARs among older adults interacting with the temi robot. As part of the visual design aspect, at the beginning of the session, some of the participants could add visual characteristics to certain robot parts, focusing on the design of the robot's head shape (Figure 10.a), apron and screen graphic (face). Then, the experimental tasks and the interaction experience, i.e., the functional and behavioral tasks were executed (for all participants). Each experimental session lasted ~45 minutes. Participants' oral and behavioral reactions were collected to build a design space model. The design space model is aimed to enable better understanding of the meaning of the robot in the eyes of older adults.

### 5.4.2. Participants

Participants were volunteers, older adults aged 70-80 from an assisted living facility "Palace Lehavim". 31 participants were recruited, 18 women and 13 men participated in the experiment, some of them living together with spouses and some living alone. Participants were recruited by creating interest in the study by strolling along the assisted living lobby along with temi who invited them to participate. In addition, messages were sent through their joint WhatsApp group, and we were approached by people who wanted to participate.

## 5.5 Results
### 5.5.1. Selected visual designs

At the beginning of the experiment, we asked some of the participants to customize the robot by their personal taste from a defined set of choices (see Figure 10). Their choices were varied. More men chose the image of the child, while more women chose the image of the dog, see Table 8. When presented with the table with all the design elements and explaining that they can design temi by them, most of the participants expressed enthusiasm and desire for this part (e.g., "What a beauty", "Wow"). Also, in many cases, they loudly expressed the intentions behind their choices. For example: "I want to design a girl/boy robot so I will choose...", "I want to design a serving robot so I will choose..."). In the initial introductory conversation with temi (see section 5.2) – they showed more interest in the conversation. This is reflected by smiles throughout the conversation and/or loud laughs after temi's joke (the joke: "You don't look over 25!"). These participants tended to create an independent conversation with temi in intermediate moments during the experiment. For example, at the end of the navigation task when the robot returns to the starting point. Some reactions and independent conversations were "Come to me", "How do you know to come to me so nicely", "Well done!". In general, the enthusiasm of the participants was felt. Their experiments passed in a less systemic way, they were interested, talked to temi beyond what was expected and were enthusiastic during the experiment.

Table 8  Robot character chosen by participants, by gender.

|       | White child head | Gold dog head | Black dog head | Yellow flowery head |
|-------|------------------|---------------|----------------|---------------------|
| Women | 1                | 4             | 1              | 1                   |
| Men   | 5                | 0             | 1              | 2                   |
| **Total** | **6**        | **4**         | **2**          | **3**               |

### 5.5.2. Difficulties

Some participants had trouble with the navigation task. For example, when a participant was unable to bring the robot to the end of the path where the image was placed, she took the tablet and began to tune the robot with the help of the arrows. The arrows disappear when prolonged contact lessness and therefore as soon as they disappear and the participant clicked on the screen on their location, the robot began to move in the direction of the pressing but without stopping until further intervention. The user didn't understand why the robot continues to advance and the experimenter had to provide her with further guidance. Later, she still couldn't direct the robot in the direction she wanted and began to show signs of frustration with the robot's inability to perform the desired actions. The participant continued to try to direct the robot and reached a point where it continued in the completely opposite direction. Other participants like her, who were unable to complete the navigation task, showed frustration throughout the rest of the experiment as well. Their frustration manifested itself in a reluctance to cooperate later in the task of cognitive activity in which they settled for a single game just to "please" the experimenter.

Other difficulties that were noticed during the experiment were robotic failures. For example, a failure to initiate the video call at the right part of the navigation task, due to a synchronization problem between the tablet and the robot, or problems with the internet connection. Although failures of this type appeared, they did not seem to affect the participants' activity to complete the mission, their subsequent cooperation or spirit during the rest of the experiment.

# 6 Discussion, Constructing the Design space model

This research presents an examination of ways to elicit needs and integrate the older adults' population in the design process of socially assistive robots. First, we conducted interviews with older adults, relatives, and caregivers to understand the demands and needs of the older adults from several different perspectives and analyzed those interviews. Then we conducted two focus groups that combined those requirements and gave the older adults a place to express themselves in the design of the robot itself as well. From analyzing those initial steps, we decided to concentrate on three aspects: functionality, behavior, and visual design that are at the core of the design space model. Afterward, we developed an experiment addressing those aspects and with the same participants, we conducted an advanced participatory focus group to use their experience with a SAR to delve further into needs and perceptions.

From the thematic analysis of the interviews, we found two main themes. Each of the themes was split into several sub-themes (see section 3.2.3). We found that there are different approaches of stakeholders to robots for the older population. There are older people who are afraid of robots and there are older people who see the robot as a lifeline (physically and mentally) and as something that can be used in case of difficulty. Similar findings were reported by others (e.g., [27]), that found that technophobia of older adults exists, yet there are also different patterns, regarding robots, depending on the level of acceptance of the participants. On the other hand, there were those who expressed a lack of need for help or reported that they find solutions for different situations that happen to them daily without defining them as problems or difficulties. Others criticized the use of robots and there were also those who found solutions for other older people who may need it, such as people who are older than them or disabled. Similar findings were described in [28], a study that explored the difficulties and needs of adults with Mild Cognitive Impairment (MCI) and their attitudes regarding the use of a robot for elder care. They found that older adults reported some difficulties in their daily lives, but those that found adapting ways to handle them, did not see themselves as in need of help. In addition, they testified that they did not need or want a robot now, and some even balked at such use. However, they did see it as effective for the future when they would need help. Interviewing three stakeholder groups enabled us to map additional needs and requirements that can arise when looking at different perspectives. As Östlund et al. [29] claim, older adults do not tend to be involved and aware, and if they are, their participation is limited [30]. We found that there are different requirements from the robot between the stakeholders' groups (see Table 6). Previously, Johnson et al. [31] found that older adults' desires from the robot are related to general help that would make them feel independent and free, while the caregivers wanted aid related to actions of health, safety, and monitoring. When we compared gender differences, the results showed differences between men and women when considering needs (Figure 2). Orji [32] showed that accepting technology has a different meaning depending on gender, and that women are less open to technology than men. We did not find that women are less open, but we did find different requirements and expectations.

Both focus groups utilizing the toolkit created robots with clear eyes or faces and with some mobile capabilities from the robot part repository presented to them (see Table 4). This is consistent with Kuo et al. [33], claiming that it's a human tendency to assign personality and characteristics to nonhuman beings. We also compared the "ideal" robot outcome of the toolkit between the two focus groups and found differences in the color selected (dark versus white and blue combination) but similarities with the parts chosen for assembling the robots (see Figure 7 and 8). In general, although we indicated differences between men and women, and between the groups of stakeholders, we found similarities between the two older adult focus groups. Still, there are differences even among groups of older adults with very similar sociodemographic background and age.

Practical experiments were constructed using different tasks and behaviors to examine each of the factors. We built a navigation task and a cognitive activity, created a dialogue that included humor between the robot and the participants, and let some of the participants design the robot by adding external elements (see Table 5; Table 6; Table 7). We found that difficulties appeared in relation to the success of the navigation task or to general difficulties of the robot. When a participant had difficulty in completing the navigation task, we found that these participants were disappointed, and as a result, the rest of their experiment was impaired. In addition, sometimes malfunctions appeared in the robot that are not related to the operation of the older adult user, but despite this, no effect was observed on the spirit of the participants and their desire to continue the experiment. Similar results were obtained in a study [34] that examined the usability of different interfaces with the aim of developing advanced robot services to provide older people with independent lives. Also though, malfunctions in our robotic system were observed, that were not related to the participant's actions, and despite that, no strong usability problems were observed because of these malfunctions.

Each experimental tool added another layer and contributed to the construction of the design space model (see Figure 11) in which we can see the connections between the factors that we examined, the abilities of the robot, and the observed emotions of the participants. Previous studies such as [35], which examined the impact of the robot's role, social skill, and appearance on user acceptance, have examined permutations of these aspects but without searching for the relationship between them. Already in conducting the interviews, we had a vague idea about the fields we wanted to approach, therefore we combined questions to sharpen those ideas into factors. With the interview questions, we examined how the characterization of the needs of older adults can be functionally expressed in a robot. From the understanding of the daily actions and the way in which older adults spend their time, we could learn about the way in which the robot should behave in an environment of older users. In terms of the exterior design of the robot, we combined questions that contributed to their understanding of existing robots and how they are perceived (see Table 1). To learn about the functionality in the focus groups, we focused on the areas that came up in the interviews and examined different contexts that meet the needs that arose in the interviews. In terms of the robot's behavior, we used the focus groups to learn from existing robots on the market about the desired behavior of the robot while integrating its functionality. Using the toolkit (see Figure 2), we were able to learn about the preferences of older adults in robot design and how the toolkit helps them express their desires and perceptions. The practical experiment that examined the three aspects separately allowed us to examine the feasibility of each of the aspects and how they can be improved and how they can help maintain the quality of life of the users. We found that there is great importance to mapping the reactions of the users in the process. Those can reveal their true feelings about the robot upfront even if these feelings are not explicitly stated.

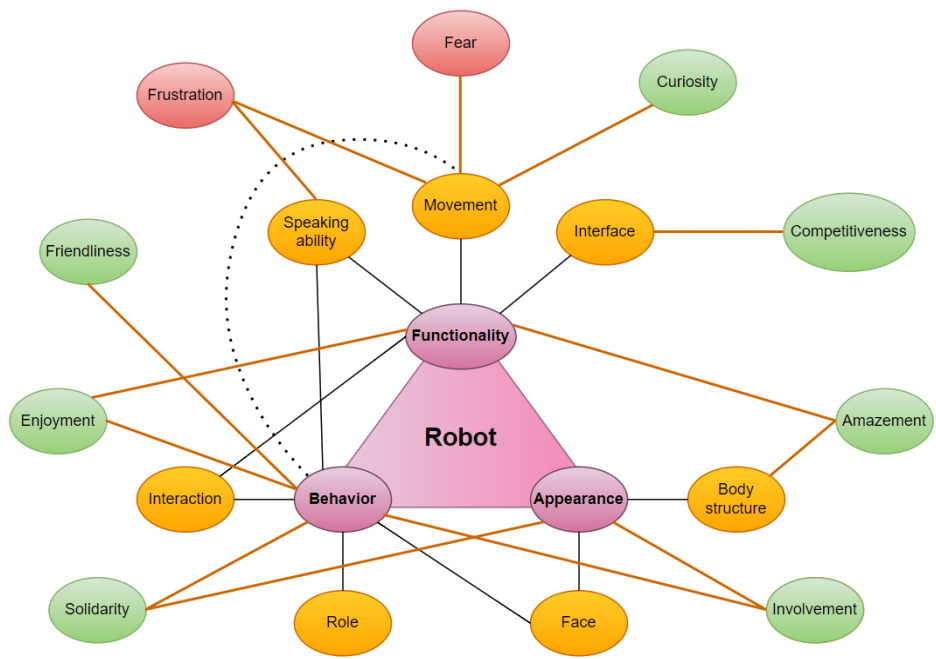

**Figure 11** The Design space model. The inner triangle consists of the factors examined in the experiment (functional role, character/behavior and appearance, the middle circle consists of the robot's abilities examined in each field, and the external circle consists of participants' observed emotions and how they are related to the interaction with the robot.

We analyzed participants' behavior throughout the experiment and conducted a qualitative analysis of the data. Starting by gathering participants' statements and then analyzing those to identify feelings and attitudes that the participants expressed toward the robot. From this data, we constructed the design space model described in Figure 11. The base of the model contains a triangle of the factors we examined in the experiments: functional, behavioral, and visual design. Around the inner triangle, we built the middle circle that represents the capabilities of the robot in each field (marked in orange). From the capabilities of the robot, we mapped the emotions expressed by the participants and linked them to the relevant field or ability of the robot (marked in red for negative emotion and green for positive). A connection to an entire factor, describes that the emotion was expressed across all the capabilities of the robot from that topic. Black lines connect the fields to the capabilities of the robot, and a dashed line represents a connection that wasn't tested in the experiment but is known as related from the literature. Orange lines connect the robot's capabilities to the participants' emotions. Table 9 provides more specific examples.

**Table 9** Description of the Design space model.

| | |
|---|---|
| **Functional:** This factor included three capabilities of the robot, the ability to speak, move and interact. | |
| **Fear** | Fear appears with the robot's ability to move. Participants experienced fear when the robot approached them but did not stop. When the participant moved from their initial position or did not understand when the robot needed to stop accurately it led to a fear response, e.g. "Isn't it going to stop?" with a physical response of moving backward. |
| **Curiosity** | Curiosity is linked to the robot's ability to move. Participants expressed curiosity about the robot at the start of the experiment when it approached them at the starting point, e.g. "Is it working on its own?" Many participants expressed these kinds of responses and were curious about the robot's movement, which is an innovative action for them that even caused giggling sometimes. |
| **Competitiveness** | This emotion is linked to the interface of the robot because for some of the participants the cognitive activity with the robot felt like an entity competing with them. E.g., "the robot is not as good as I am" with a smile of defeating the opponent. |
| **Frustration/ disappointment** | Frustration is linked to speech and movement abilities. The robot has a robotic voice, so it could not correctly pronounce some names of the participants. Some participants expressed disappointment when it did not pronounce their name correctly, e.g., "It doesn't sound good that it doesn't pronounce my name correctly". There were also those who were frustrated for not being able to control its movement remotely and therefore seemed frustrated for the duration of the experiment starting from that unsuccessful part. |
| **Visual Design:** The element of the inner triangle associated with this area is called "Appearance". This area included two capabilities of the robot, body structure, and face. | |
| **Amazed** | A sense of amazement was found in the early stages of the experiment when participants were asked to choose visual design characteristics for the robot, e.g., "Wow, what a beauty" at the sight of the colors and elements placed on the table. When we tested the robot's functionality, we also discovered reactions of astonishment towards the robot's capabilities, the way it moves in space, how it make a conversation, and what we presented in the interface. |
| **Involvement** | The body language of the participants and their desire to take part in the experiment were felt at the design part and also during the conversations with the robot. We incorporated humor into the conversations with the robot and found that in most cases participants laughed. In addition, they took an active part in the conversation and waited for feedback. In the design part, the involvement was reflected when the participants did not randomly select the elements but gave an explanation about the meaning of the choice and the intention behind it. |
| **Behavioral:** This area included two capabilities of the robot, interaction, and role. | |
| **Solidarity** | Many of the participants saw the robot as a human-like figure and therefore humanized it in the discussion they had, e.g., "What am I going to talk to a dog" when the participant looked at an element of a dog head and finally chose the element of a child head. Other responses included cheers when the robot approached as if encouraging a little child to perform an action or giving attribution, such as "Sir". In addition, humanity was also found in the reactions to the appearance of the robot. Some participants tried to create a character of a butler, a boy, or a girl using the design elements that were at their disposal. |
| **Enjoyment** | We found that the participants benefited from the robot's functionality as well as the robot's behavior. The participants tended to comment that the games are fun and that they enjoy them, e.g. "I really enjoyed playing with the robot". In addition, we noticed that the connection with the robot in conversations led to laughter, smiles, and enthusiasm. |
| **Friendliness** | We noticed that many participants saw the robot as a sociable entity, and this manifested itself as a response to the robot's behavior, e.g., "He's my friend", "Come get close to me," and more. |

# 7   Conclusion, Limitations and Future work

This study included interviews, focus groups and a practical one-on-one HRI experiment to produce methods, requirements and understanding of the emotions that participants had in the experience with the robot. We found gaps between different stakeholders, among the older participants and between genders. Therefore, it can be concluded that it is important to listen to varied populations and not focus on just one group. In addition, we can argue that older people have strong opinions about what they need. Furthermore, there are differences in users' perceptions. We found that the general requirements are relatively the same, but the way people perceive each demand and how they want it to be implemented varies.

Our study has some limitations. In the interview and exploratory focus groups we sampled older adults that live alone in the southern region of Israel. In the experimental stage, we collaborated with "Palace Lehavim", an assistive living facility, in the same region. Yet, our sample is only a segment of the population of the southern region of Israel. Participants who volunteered to participate in the first stage represent a sample of convenience and are all from higher socioeconomical statuses. Due to COVID-19, it was difficult to form focus groups with older adults. On the other hand, the need to maintain social distancing and conduct the study during COVID-19 may have affected participants' considerations, expectations, and requirements from a SAR. When considering the toolkit, the design elements presented to the participants were limited, they were not given infinite options to choose from. Different elements or more elements may have led to different results. Furthermore, the formation of the design space model was constructed by qualitative analysis and not by utilizing advanced quantitative analysis tools such as voice analysis or technological devices.

Future work should include additional studies that will examine improved or additional relationships of human-robots beyond friendship. In the future, it will be worthwhile to develop more active behaviors of the robot to ensure a more advanced relationship with a robot that can learn about the user and perform actions accordingly. Future planning should address the small details that can affect planning for the older adult population, such as making the product accessible to older adults. It is possible to expand the design space model, examine each emotion separately or use advanced technological tools to improve the mapping.

# References

[1] R. M. Johansson-Pajala, K. Thommes, J. A. Hoppe, O. Tuisku, L. Hennala, S. Pekkarinen, H. Melkas and C. Gustafsson, "Care robot orientation: what, who and how? Potential users' perceptions," *International Journal of Social Robotics,,* pp. 1-15, 2020.

[2] "Schulz, R., Eden, J., & National Academies of Sciences, Engineering, and Medicine. (2016). Older adults who need caregiving and the family caregivers who help them. In Families caring for an aging America. National Academies Press (US).".

[3] "Glende, S., Conrad, I., Krezdorn, L., Klemcke, S., & Krätzel, C. (2016). Increasing the acceptance of assistive robots for older people through marketing strategies based on stakeholder needs. International Journal of Social Robotics, 8(3), 355-369.".

[4] S. Cooper, Ó. Villacañas, L. Marchionni and F. Ferro, "Robot to support older people to live independently," *arXiv preprint arXiv:2104.07799,* 2021.

[5] W. A. Rogers, T. Kadylak and M. A. Bayles, "Maximizing the Benefits of Participatory Design for Human–Robot Interaction Research With Older Adults," *Human Factors, 64(3),* pp. 441-450, 2022.

[6] "Pelau, C., Dabija, D. C., & Ene, I. (2021). What makes an AI device human-like? The role of interaction quality, empathy and perceived psychological anthropomorphic characteristics in the acceptance of artificial intelligence in the service industry. Comp".

[7] I.-H. Kuo, C. Jayawardena, E. Broadbent and B. A. MacDonald, "Multidisciplinary design approach for implementation of interactive services," *International Journal of Social Robotics Kuo, I. H., Jayawardena, C., Broadbent, E., & MacDonald, B. A.,* pp. 3(4), 443-456, 2011.

[8] E. Liberman-Pincu and T. Oron-Gilad, "Exploring the Effect of Mass Customization on User Acceptance of Socially Assistive Robots (SARs)," *In Proceedings of the 2022 ACM/IEEE International Conference on Human-Robot Interaction,* pp. (pp. 880-884), 2022, March.

[9] E. Broadbent, R. Stafford and B. MacDonald, "Acceptance of healthcare robots for the older population: review and future directions," *International journal of social robotics, 1(4),* pp. 319-330, 2009.

[10] "S. A. Olatunji, T. Oron-Gilad, N. Markfeld, D. Gutman, V. Sarne-Fleischmann and Y. Edan, "Levels of Automation and Transparency: Interaction Design Considerations in Assistive Robots for Older Adults," in IEEE Transactions on Human-Machine Systems, vol. 5".

[11] "Olatunji, Samuel, Oron-Gilad, Tal, Sarne-Fleischmann, Vardit and Edan, Yael. "User-centered feedback design in person-following robots for older adults" Paladyn, Journal of Behavioral Robotics, vol. 11, no. 1, 2020, pp. 86-103. https://doi.org/10.1515/pjb".

[12] C. R. Wilkinson and A. De Angeli, "Applying user centred and participatory design approaches to commercial product development," *Design Studies, 35(6),* pp. 614-631, 2014.

[13] "Afrianto, I., & Atin, S. (2018). The Journal Aggregator System Concept Using User Centered Design (UCD) Approach. IJNMT (International Journal of New Media Technology), 5(2), 71-75.".

[14] "Eftring, H., & Frennert, S. (2016). Designing a social and assistive robot for seniors. Zeitschrift für Gerontologie und Geriatrie, 49(4), 274-281.".

[15] "Rogers, W. A., Kadylak, T., & Bayles, M. A. (2021). Maximizing the Benefits of Participatory Design for Human–Robot Interaction Research With Older Adults. Human Factors, 00187208211037465.".

[16] M. Yasuoka-Jensen and T. Kamihira, "How participation is practiced?–Extension of Participatory Design Model. In Service Design Geographies," *Proceedings of the ServDes. 2016 Conference,* vol. Linköping University Electronic Pre, pp. No. 125, pp. 279-291, 2016.

[17] "Bardaro, G., Antonini, A., & Motta, E. (2021). Robots for elderly care in the home: A landscape analysis and co-design toolkit. International Journal of Social Robotics, 1-25.".

[18] "Kuo, I. H., Jayawardena, C., Broadbent, E., & MacDonald, B. A. (2011). Multidisciplinary design approach for implementation of interactive services. International Journal of Social Robotics, 3(4), 443-456.".

[19] "Pnevmatikos, D., Christodoulou, P., & Fachantidis, N. (2021). Designing a Socially Assistive Robot for Education Through a Participatory Design Approach: Pivotal Principles for the Developers. International Journal of Social Robotics, 1-26.".

[20] "Johnson, M. J., Johnson, M. A., Sefcik, J. S., Cacchione, P. Z., Mucchiani, C., Lau, T., & Yim, M. (2020). Task and design requirements for an affordable mobile service robot for elder care in an all-inclusive care for elders assisted-living setting. Inte".

[21] "Bradwell, H. L., Noury, G. E. A., Edwards, K. J., Winnington, R., Thill, S., & Jones, R. B. (2021). Design recommendations for socially assistive robots for health and social care based on a large scale analysis of stakeholder positions: Social robot desi".

[22] "Lee, H. R., Šabanović, S., Chang, W. L., Nagata, S., Piatt, J., Bennett, C., & Hakken, D. (2017, March). Steps toward participatory design of social robots: mutual learning with older adults with depression. In Proceedings of the 2017 ACM/IEEE internation".

[23] V. Clarke, V. Braun, G. Terry and H. N., "Thematic analysis," in *Handbook of research methods in health and social sciences*, 2019, pp. (pp. 843-860).

[24] V. Braun and V. Clarke, Thematic analysis2012 ,.

[25] C. McNaught and P. Lam, "Using Wordle as a supplementary research tool," *Qualitative Report, 15(3),* pp. 630-643, 2010.

[26] E. Liberman-Pincu, Y. Parmet and T. Oron-Gilad, "Judging a socially assistive robot (SAR) by its cover; The effect of body structure, outline, and color on users' perception," *arXiv preprint arXiv:2202.07614,* 2022.

[27] O. Zafrani, "Between Fear and Trust: Factors Influencing Older Adults' Evaluation of Socially Assistive Robots. ," *arXiv preprint arXiv:2207.05387.,* 2022.

[28] Y.-H. Wu, V. Cristancho-Lacroix, C. Fassert, V. Faucounau, J. de Rotrou and A.-S. Rigaud, "The attitudes and perceptions of older adults with mild cognitive impairment toward an assistive robot," *Journal of Applied Gerontology, 35(1),* pp. 3-17, 2016.

[29] B. Östlund, E. Olander, O. Jonsson and S. Frennert, "STS-inspired design to meet the challenges of modern aging. Welfare technology as a tool to promote user driven innovations or another way to keep older users hostage?," *Technological forecasting and social change, 93,* pp. 82-90, 2015.

[30] S. Frennert and B. Östlund, "Seven matters of concern of social robots and older people," *International Journal of Social Robotics, 6(2),* pp. 299-310, 2014.

[31] M. J. Johnson, M. A. Johnson, J. S. Sefcik, P. Z. Cacchione, C. Mucchiani, T. Lau and M. Yim, "Task and design requirements for an affordable mobile service robot for elder care in an all-inclusive care for elders assisted-living setting," *International journal of social robotics, 12(5),* pp. 989-1008, 2020.

[32] R. O. Orji, "Impact of gender and nationality on acceptance of a digital library: An empirical validation of nationality based UTAUT using SEM," *Journal of Emerging Trends in Computing and Information Sciences, 1(2),* 2010.

[33] I. H. Kuo, J. M. Rabindran, E. Broadbent, Y. I. Lee, N. Kerse, R. M. Q. Stafford and B. A. MacDonald, "Age and gender factors in user acceptance of healthcare robots," *In RO-MAN 2009-The 18th IEEE International Symposium on Robot and Human Interactive Communication (pp. 214-219). IEEE,* 2009, September.

[34] "Bevilacqua, R., Felici, E., Marcellini, F., Glende, S., Klemcke, S., Conrad, I., ... & Dario, P. (2015, August). Robot-era project: Preliminary results on the system usability. In International conference of design, user experience, and usability (pp. 553".

[35] I. A. Hameed, Z. H. Tan, N. B. Thomsen and X. Duan, "User acceptance of social robots," *In Proceedings of the Ninth international conference on advances in computer-human interactions (ACHI 2016), Venice, Italy (pp. 274-279),* 2016, April.

[36] R. Schulz, J. Eden and a. M. National Academies of Sciences Engineering, "Older adults who need caregiving and the family caregivers who help them," *In Families caring for an aging America,* p. National Academies Press (US), 2016.

[37] S. Glende, I. Conrad, L. Krezdorn, S. Klemcke and C. Krätzel, "Increasing the acceptance of assistive robots for older people through marketing strategies based on stakeholder needs," *International Journal of Social Robotics, 8(3),* pp. 355-369, 2016.

[38] C. Pelau, D.-C. Dabij and I. Ene, "What makes an AI device human-like? The role of interaction quality, empathy and perceived psychological anthropomorphic characteristics in the acceptance of artificial intelligence in the service industry," *Computers in Human Behavior, 122,* p. 106855, 2021.

[39] M. Axelsson, R. Oliveira, M. Racca and v. Kyrki, "Social Robot Co-Design Canvases: A Participatory Design Framework," *ACM Transactions on Human-Robot Interaction (THRI),* pp. 11(1), 1-39, 2021.

[40] O. Zafrani, T. B. White and H. Riemer, "When your favorites disappoint: Self-construal influences response to disappointing brand experiences," *Current Psychology,* pp. 1-12, 2021.

[41] "Johansson-Pajala, R. M., Thommes, K., Hoppe, J. A., Tuisku, O., Hennala, L., Pekkarinen, S., ... & Gustafsson, C. (2020). Care robot orientation: what, who and how? Potential users' perceptions. International Journal of Social Robotics, 1-15.".

[42] "Broadbent, E., Stafford, R., & MacDonald, B. (2009). Acceptance of healthcare robots for the older population: review and future directions. International journal of social robotics, 1(4), 319-330.".

[43] "Liberman-Pincu, E., Van Grondelle, E. D., & Oron-Gilad, T. (2021, March). Designing robots with relationships in mind: Suggesting two models of human-socially assistive robot (SAR) relationship. In Companion of the 2021 ACM/IEEE International Conference o".

[44] "Liberman-Pincu, E., & Oron-Gilad, T. (2022, March). Exploring the Effect of Mass Customization on User Acceptance of Socially Assistive Robots (SARs). In Proceedings of the 2022 ACM/IEEE International Conference on Human-Robot Interaction (pp. 880-884).".

[45] "Cooper, S., Villacañas, Ó., Marchionni, L., & Ferro, F. (2021). Robot to support older people to live independently. arXiv preprint arXiv:2104.07799.".

# Appendix A. Daily needs of older adults who live alone

Table 10 and Table 11 provides the themes that emerged from the interviews in the first part of the research.

**Table 10** Daily needs of older adults living alone, the first main theme.

| Theme: The daily needs of older adults who live alone | | |
|---|---|---|
| **Subtheme: Needs** | **Subtheme: Unresolved needs** | **Subtheme: Solutions the older adults found** |
| "I usually like watching TV, I have my shows, and my sister and sister-in-law calling me so I talk to them a lot." | "I fell in the shower, but there's no way I'm not taking a shower, I live with it." | "My children help me to go up the stairs, they wait with me until I succeed to go up." |
| "I love being with people and on Saturday I go to my kids and see the entire family." | "I need someone to talk to me so I can forget about my blood pressure until it's down." | "When my blood pressure is dropping, I try to take more pills to drink tea with lemon, but if that doesn't help, I'm going to bed." |
| "I go to gym classes to maintain a healthy lifestyle. I have an iPad and I play it and also on the computer." | "I don't want any caretaker, I don't trust them, a robot doesn't steal from you." | "I fill my day with things to do so that I won't feel lonely." |
| "I have a lot of interaction with friends in the assisted living facility, after lunch we sit and talk." | "Sometimes I don't understand the bank on the phone or all sort of entities that have switched to give service on the phone." | "I have a diary and every night before bed I write my chores for the next day." |
| "I'm missing the feelings of warmth and love." | | "I get a newspaper every Sunday and every Friday." |
| "I'm passing the time around with food. I like to cook to pass the time." | | "If something is hard for me, I just don't do it, nobody can make me to do it." |
| "When we were young, we used to play and do things alone, we used to go outside and not sit at home all day. We need it." | | |
| "When I was young, people played games outside, too bad today it's not like that." | | |
| "I need someone to hug me." | | |
| **Subtheme: Needs from the past** | **Subtheme: Needs following medical treatment** | **Subtheme: Lack of the need to get help** |
| "I used to go to a daily care where they would give us physiotherapy and there were classes and I really enjoyed it." | "I should always check that my sugar level is fine and know what's going on with my blood pressure." | "I don't need help because I'm still not that old, I do things by myself." |
| "I love that my sister-in-law comes to sit with me in the afternoon, but she can't come anymore." | "I get up from my nap at 3 pm and make myself tea with cake to balance my sugar level." | "You have to do things on your own, don't need an assistive robot." |
| "I was very active in the building, I used to clean the whole building, now both the old age and the surgery won't let me. I'd like to be more active." | "Because I've had a knee surgery, I need to walk around to strengthen my leg." | "When I cook, I look at the recipe and have to rest occasionally because of the waiting time in the recipe so I rest in the preparation and then I'm fine." |
| "I feel like I'm doing things, but I'm not the same person as I used to be." | "Because of my illness, I'm not allowed to be in the sun, so I try to leave the house only when it's early, in the winter or when the sun comes down." | "I don't need a robot I do things alone. It'll just get in the way. I also don't want it to talk to me because it will irritate me." |
| "I really like to hang out and shop but I can't anymore. I have iron deficiency, and it weakens me." | "I'm taking pills so I have to eat because I have hypertension." | "I shower alone and dress alone, even though it's hard, I try to do it myself." |
| "I used to go out to the theater." | | "I clean once a week, there's no reason to clean more than that because people barely come here anyway." |
| | | "I would like to go out at noon but it doesn't matter to me I'm alone and I'm used to being alone." |

**Table 11** Attitude towards robots, the second main theme

| Theme: Attitude towards robots | | |
|---|---|---|
| **Subtheme: Fears from robots** | **Subtheme: The fears robots should moderate** | **Subtheme: Criticism of the use of technology** |
| "Yes, the robot scares me because it can cause degeneration, it makes a man lazy." | "If there's a robot here, I'll have someone who's always here and keep me from falling." | "If there will be a robot, I'll get used to doing nothing. So, a robot isn't so good." |
| "I wouldn't want that. I'm afraid of robots, it's dangerous. Dangerous for the health, people need to do things alone." | "Knowing that if I forget something, I could send the robot to do it." | "What do I need a robot for? you have to do things on your own." |
| "Depending on which robot, a robot doing things at home is fine but not robots who replace humans like they make in the world, I'm very afraid of it." | "I used to ride a bike, now I'm scared because I've grown up and I can suddenly fall and there's no one to be with me in this." | "It's great the fields technology progress today." |
| "The robot can break down and then I won't know what to do with it so I just won't use it." | "If it can help health, that's what I need." | |
| "I've seen movies where the robots have taken over and I'm afraid of it." | "It needs to smile all the time and to give a good mood." | |
| "From a few movies I've seen, I don't want them to get mad at me and I also don't want it to break things." | "It needs to monitor my parent and update me that he has taken his medications and that everything is fine with him." | |
| "If the robot will help me outside the house, it can cause panic to the people around." | "I need to know if my mother fell in the house." | |

| **Subtheme: Finding solutions for others or the future** | **Subtheme: Difficulties** |
|---|---|
| "My vision is weakening, so it's hard for me to read books, but I really like to read." | "When I will be older, I think I would need help to get up." |
| "I'd like the robot to be able to help me bend over. For example, when I arrange the food in the refrigerator, it is difficult for me to bend to the low shelves or to put my clothes in the closet. Generally, in all of the physical tasks I could use some help." | "I think that in older age I'd like a robot to remind me where things are." |
| "I don't do walks anymore because I'm more tired so I need to rest. I'm getting lazy to get out." | "There are people in a wheelchair or older people who may need more help I think I'm fine." |
| "I like to go to the pool by foot, but it's harder for me after the surgery." | "Maybe helping out in the shower, being able to help to get in there or help getting dressed but it's not for me I'm trying to think about what other people who need help want." |
| "When I'm making lunch preparing the food can be a little difficult to stand for so much time." | |
| "Because I broke my shoulder bone it's harder to cut vegetables and meat." | |
| "I'm alone here at home and most of the time I spend alone so there can be some feeling of loneliness." | |
| "It's hard for me to get dressed and I don't want to go out in pajamas, so I stay at home a lot." | |
| "On sick days I have sometimes difficulty in standing in the shower, but I don't give up." | |
| "I'm missing my children; the vast majority don't live in Beer-Sheva." | |
| "I used to go out to the theater." | |

# Appendix B. Daily routine process

Another outcome of the interviews was the creation of daily-routine charts of older adults. Map the daily routine of older adults, through the questions asked in the interviews, gave us better understanding of how robots should behave around older adults during the day. By understanding what older adults' experiences during the day, we could identify places where it will be possible to help older adults. Figure 12 shows an example of a daily routine from one of the interviewees. In the routine presented in Figure 12, one can see that the interviewee starts the day with a morning routine that includes taking medication and breakfast, then talking to friends, watching videos of the family on WhatsApp, and watching TV. Later, the interviewee eats lunch and is available for recreational activities of sewing. Throughout the interview, the interviewee noted that she talks to people and communicates with the outside world through voice communication. The interviewee's need to communicate with her surroundings, avoid silence through the TV or through speech stood out throughout this interview.

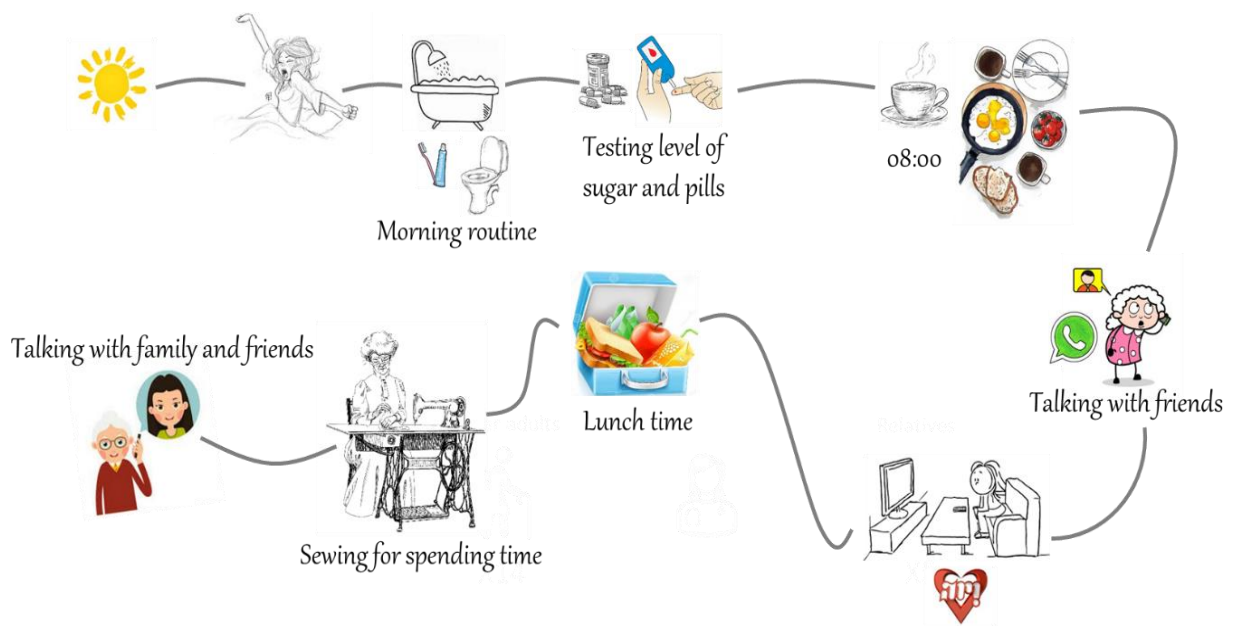

**Figure 12**  Example of one daily routine, as derived from the structured interviews.